\documentclass[a4paper,showkeys,floatfix,aps,pre,reprint,groupedaddress]{revtex4-1}
\usepackage{amsmath,amssymb}
\usepackage{graphics,graphicx}
\usepackage{dcolumn,bm}
\usepackage{psfrag}
\usepackage{soul}
\usepackage{enumitem}

\begin{document}

\title{Modes of failure in disordered solids}

\author{Subhadeep Roy${}^{1,2}$}
\email{sroy@eri.u-tokyo.ac.jp}
\author{Soumyajyoti Biswas${}^{3}$}
\email{soumyajyoti.biswas@ds.mpg.de}
\author{Purusattam Ray${}^{1}$}
\email{ray@imsc.res.in}
\affiliation{
${}^1$ The Institute of Mathematical Sciences, Taramani, Chennai-600113, India.\\
${}^2$ Earthquake Research Institute, University of Tokyo, 1-1-1 Yayoi, Bunkyo, 113-0032 Tokyo, Japan.\\
${}^3$ Max Planck Institute for Dynamics and Self-Organization, Am Fassberg 17, G\"{o}ttingen, Germany. 
}

\date{\today}

\begin{abstract}
\noindent 
The two principal ingredients determining the failure modes of disordered solids are the strength of heterogeneity and the length scale of the region affected in the solid following a local failure. While the latter facilitates damage nucleation, the former leads to diffused damage -- the two extreme nature of the failure modes. In this study, using  the random fiber bundle model as a prototype for disorder solids, we classify all  failure modes that are the results of interplay between these two effects. We obtain scaling criteria for the different modes and propose a general phase diagram that provides a framework for understanding previous theoretical and experimental attempts of interpolation between these modes. As the fiber bundle model is a long standing model for interpreting various features of stressed disordered solids, the general phase diagram can serve as a guiding principle in anticipating the responses of disordered solids in general.
\end{abstract}


\maketitle


\section{Introduction}
Response of a disordered solid subjected to stress provide a vital route in predicting imminent breakdown in those systems. Understanding such responses is a major goal for myriads of situations starting from micro-fracture  to earthquakes  \cite{books}. The apparent independence of the effect of the structural details in the static and dynamic responses of the disordered solids, for example roughness of a fractured front, avalanche size distributions etc., fueled decades of efforts in modeling these phenomena using simple, generic and minimal ingredients \cite{rmp1}. The focus of these studies is on the understanding of the mechanical stability of the systems, precursor to catastrophic failure and also to explore the possibility of universality of the above mentioned response statistics in the sense of critical phenomena.  However, while there can be scale free behavior of response functions indicating criticality in some cases, there can also be nucleation driven abrupt failures in others. Therefore, such association of fracture with critical phenomena is not straight forward (see e.g. \cite{moreno, zapperi97}). 

It is, however, known that the two main factors that determine such modes of failures are the strength of disorder and the range of interaction within the solid in terms of stress transfer. The aim of this work is to classify all the phases arising out of the interplay of these two effects and to arrive at criteria in distinguishing  such phases, thereby providing a framework for understanding all the modes of failure using a simple model for the disordered solids.

It is known experimentally that the presence of heterogeneity increases the precursory signals prior to failure \cite{vass}. The strain energy is dissipated within a short range of crack propagation in heterogeneous solids, as opposed to those lacking heterogeneity. Strong heterogeneities, therefore, compel the system transit from a brittle like to a quasi-brittle like failure mode \cite{wong}.  Such a transition in porous media was observed in Ref. \cite{Li92}, while the disorder (porosity) spanned two decades in magnitude.  The apparent contradiction of scale free size distribution for acoustic emission and subsequent damage nucleation was also observed in Ref. \cite{guarino98}. While experimentally it is not easy to tune the strength of disorder precisely, heat treatment can tune the length scale of disorder in phase-separated glasses \cite{dalmas08}. There have been many other experiments and simulations describing the effect of increased disorder on roughness \cite{vvs06}, pattern formation in spring networks \cite{mala07,date96}, damage nucleation and percolation in random fuse models \cite{nukala04, kahng88} etc. 

As for the range of stress redistribution, linear elastic fracture mechanics predict a $1/r^2$ type load redistribution around an Inglis crack \cite{sch95, rama97}. However this form is not always guaranteed and can change due to finite width of the sample \cite{hutch}, correlation in disorder \cite{stanley}, size of agglomerate \cite{kendall} and so on. Here we attempt  to characterize the formation of spatial and temporal correlation arising out of the interplay of the stress redistribution, which enhances damage nucleation, and the presence of disorder, which leads to diffused damage \cite{curtin, amitrano, tommasi}.  

In this work, we report a phase diagram in the stress redistribution range and strength of disorder that captures all failure modes arising out of the interplay between these two. We consider the fiber bundle model \cite{rmp1,Book}, which has been widely used as a generic model for fracture in disordered system over many years. 
Among the many modeling approaches that attempt to capture the statistics of failure of disordered solids, fiber bundle model is arguably the simplest. Introduced in the textile engineering \cite{first}, it has been proven very useful in reproducing  behaviors near failure \cite{rmp1}. The avalanche statistics and also the roughness of fracture propagation front arising out of its intermittent dynamics, compares favorably with experiments \cite{bonamy09}. The model is a set of elements arranged in a lattice, each having a finite failure threshold drawn randomly from a distribution. On application of load, the elements ---fibers --- fail irreversibly and redistribute their load in a pre-defined neighborhood.

The two main ingredients of the model are the aforementioned neighborhood of load redistribution and the strength of the disorder in the failure thresholds of the individual
fibers. The two extreme ways of defining the neighborhood are the equal and local load sharing models. In the former, the load of a broken fiber is shared equally among all the
remaining intact fibers and in the latter, it is shared only with its nearest surviving neighbors. Neither of these two extremes are realistic, however they are important in establishing
limiting behaviors of the model. Particularly, for local load sharing, the local stress concentration around damage is so high that the failure statistics is governed by
extreme statistics \cite{harlow78,harlow91,phoenix74,phoenix75} and the critical load for the system decreases with system size \cite{harlow85,langer03}. On the other hand, a global load sharing model gives a finite failure threshold,
as the stress concentration is much lower here. Other than the two extremes, there has been a lot of studies that attempt to capture a more realistic way of redistributing the stress.
An obvious candidate was power-law load sharing \cite{hidalgo}, where the exponent of the power law determines the localization of stress, which we will discuss later. Among other
more realistic attempts was the one by Hedgepeth and Van Dyke \cite{hed}, applied for polymer matrix composites. 
While there can be situations such as plastic deformation, highly non-linear effects nea the crack-tip in
the above examples, where a simple redistribution rule is no longer valid, here we limit ourselves to the smooth asymptotics described by a power law
load sharing.
The asymptotic form of the Hedgepeth load sharing rule, however, is inverse cubic for two-dimensions \cite{gupta}. More detailed load sharing rules include those proposed by Okabe et al. among others \cite{okabe01,okabe02,mishra17,gupta17}. Particularly, as plastic and 
interfacial damages are considered, the load sharing for these cases interpolate between global and Hedgepeth load sharing. Furthermore, there are time dependent load sharing rules \cite{newman01,phoenix09} that also interpolate between local and global load sharing. Therefore, a substantial literature in physics and
engineering community has been developed in addressing the question of load redistribution range and their effect on stress localization and ultimately the failure threshold
of the materials, using the fiber bundle model. 

On the other hand, the disorder in the model comes from the distribution of the failure threshold.  The properties of the distribution function can influence the stress localization, and that, in turn, can determine the failure strength of the system. The 
spread of damage and the crackling noise, which can be used as a pre-cursor to catastrophic failure, is significantly affected by the presence of disorder. Particularly, 
higher disorder increases the pre-cursory events in the solids \cite{vass}. Due to its importance, there have been many efforts in looking for the effect of disorder strength  was made
on the fiber bundle model \cite{kun16,kun17}. Particularly in the global load sharing case, the effect of high disorder in known to bring the system from brittle to quasi-brittle state \cite{epl15}.
  
Using the simplicity and flexibility of the fiber bundle model, we can tune both the strength of disorder and stress redistribution range and obtain the different phases 
of failure in the fiber bundle model by varying the range of stress redistribution and strength of disorder.
 With the help of the phase diagram, we can now identify all its modes of failure, classify previous attempts to interpolate between some of those modes and most importantly arrive at scaling prescriptions in categorizing and predicting such failure modes.  The scaling prescriptions differ from their equilibrium, and often intuitive, counterparts (say, in Ising model), making them interesting also from the point of view of critical phenomena.  


\section{MODEL \& SIMULATION}

\begin{figure}[tbh]
\centering
\includegraphics[width=8.5cm, keepaspectratio]{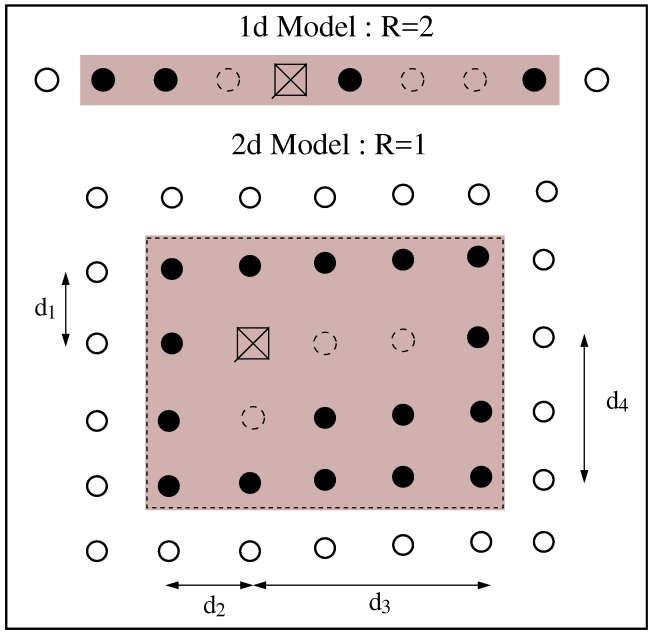}
\caption{The load redistribution region for a finite range $R$ is shown for the one dimensional and two dimensional version of the model. The intact fibers in the shaded region 
 (denoted by filled circles) are affected by the load redistribution following a failure of a fiber (denoted by a cross), while the empty dotted circles are broken fibers and 
empty circles outside the region are sites of fibers that are not affected by this event. For a more general power law load redistribution (not shown), however, all intact fibers 
are affected  but the shared load varies inversely with the distance from the broken fiber. }
\label{redist}
\end{figure}

Here we simulate the failure in fiber bundle model in one and two dimensions -- the one dimensional case is an idealized but the simplest one, while the two dimensional case is more realistic and has been used to model failure in fibrous materials (e.g. fiber reinforced composites) for many years \cite{harlow78,harlow91,phoenix74,phoenix75}.  We choose the failure thresholds of the fibers from a distribution of the form $p(x)\sim 1/x$ within a range $[10^{-\beta}:10^{\beta}]$. For high values of $\beta$, the distribution becomes very broad, making the system a highly disordered one. Physically, this implies varying strength of impurities in the system, that can significantly influence the overall critical strength of the system. Following the failure of a fiber, the load  on the failed fiber is redistributed uniformly up to a distance $R$. In one dimension, this is simply $R$ surviving neighboring fibers on either side of the failed one. In two-dimensions we search along positive and negative $x$ and $y$ axes and go up to a distance $x_+, x_-,y_+$ and $y_-$ until $R$ surviving neighbors are found along each direction (see Fig. \ref{redist}). We then redistribute the load within the rectangular region $(x_+,y_+)$, $(x_-,y_+)$, $(x_-,y_-)$, $(x_+,y_-)$ (assuming the origin at the failed fiber). Of course, there can be other choices, for example a circular region of radius $R$. While that could work well for higher values of $R$, but for smaller values there could be situations where there were no surviving fibers within that region. Moreover, such details are unlikely to affect the scaling behavior, which is also evident from the fact that our prediction matches well with power-law load redistribution studied in Ref. \cite{hidalgo}. 

With changes in these two parameters ($\beta$ and $R$) we get the different failures modes of the model. 
We will first describe the phase diagram to explain the different modes. Subsequently we will discuss the methods of drawing the boundaries and relate them to previous numerical and experimental attempts of interpolations. 



\section{Numerical Results}
Numerical results are produced for different system sizes over a wide range disorder and stress release range. Six different regions are observed through numerical simulations with individual modes of failure. 


\subsection{The $R-\beta$ Plane}
Intuitively, we expect a nucleating failure for low values of $R$ and $\beta$. This resembles brittle failures of perfectly crystalline structures. The failure thresholds of each part of the system are almost same, therefore an initial failure and subsequent load concentration around it (due to low $R$ values) compels the subsequent damages to be near that initial damage and it will continue to  grow.
\begin{figure}[tbh]
\centering
\includegraphics[width=8.5cm, keepaspectratio]{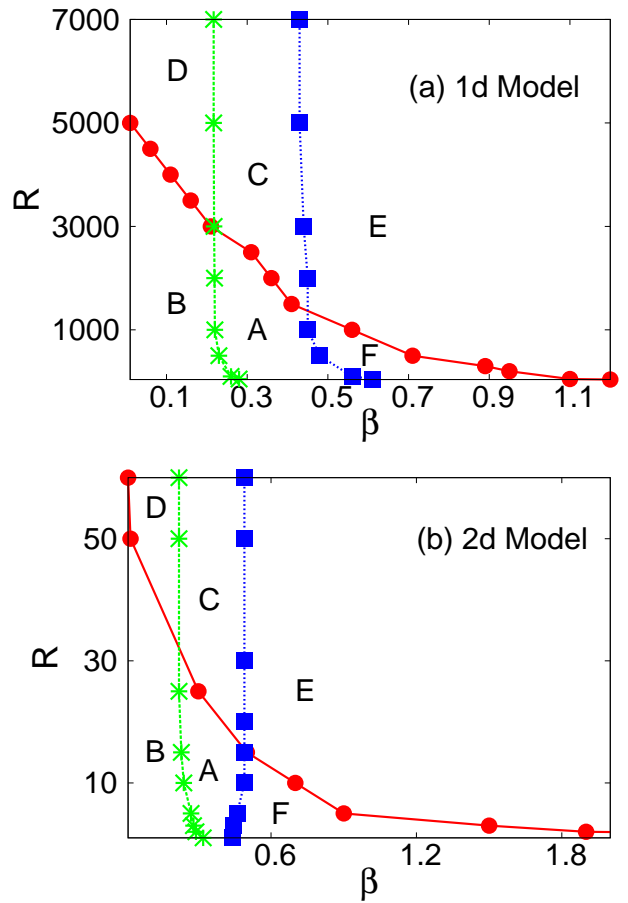}
\caption{The figure shows all the regions on $R-\beta$ plane for: (a) 1d and (b) 2d bundle. $B$ and $D$ are brittle regions and show abrupt failure. $A$ and $C$ show quasi-brittle response. Only difference is, in region $A$ and $B$ rupture process is spatially correlated. In region $E$ (spatially uncorrelated) and $F$ (spatially correlated), the failure process is mainly dominated by stress increment. }
\label{PD}
\end{figure}
Thus small $R$ and $\beta$ imply high spatial correlation in damage. This damage nucleation can be prevented by either redistributing the load of a failed fiber   to a relatively large distance, or by increasing the disorder such that the nearby fiber can have high failure threshold which compels distant fibers to fail first. 
 
 On the other hand, higher the number of fibers breaking due to stress 
redistribution, higher is the temporal correlation (we will present quantitative measures later). The temporal correlation in damage, i.e. avalanches, also behave similarly with $R$ and $\beta$. Small $R$ and $\beta$ imply higher correlation. The difference is that the temporal correlation does not vanish at the same values of $R$ and $\beta$, as the spatial correlation. The phase diagram (Fig. \ref{PD}), therefore, has regions where temporal correlation exists without spatial correlation, hence giving interesting phases for the model. 


\subsection{Description of the Phases}
Below we first describe each of the phases depicted in Fig.\ref{PD} and then describe the quantitative measures for drawing the boundaries between the phases.


\subsubsection{B: Brittle-nucleating}
In this region, as soon as the weakest fiber is broken, the entire system collapses starting from damage nucleation happening next
to the failed fiber. This is a brittle like failure (like in ceramics, say) and have both temporal and spatial correlations. The avalanche is a 
catastrophic failure here, with size $\sim L$.

\subsubsection{D: Brittle-percolating}
The system here also collapses following the breaking of the weakest fiber, but as $R$ is large enough, the subsequent damage is spatially uncorrelated i.e. multiple damage nucleation zones are formed. 

\subsubsection{A: Quasi-brittle nucleating}
In this region, the system fails after multiple stable states, hence the nature of failure is quasi-brittle. In this region, an apparent random failure eventually forms a spatially correlated failure i.e. the system begins with a scale free avalanche distribution, but for larger systems the final failure is nucleation driven (see reference \cite{Moreira,Shekhawat} for electrical analogue). 

\subsubsection{C: Quasi-brittle percolating}
This is the region where the $R$ and $\beta$ combination is such that although the spatially correlation has vanished, the temporal correlation exists. This is the region with scale free size distribution (exponent $-5/2$ \cite{rmp1}) of the avalanches. 

\subsubsection{E: High disorder limit}
In this region, neither the spatial correlation nor the temporal correlation exists. As can be seen, this region appears even for very low $R$ values, given the disorder distribution is broad enough (high $\beta$).

\subsubsection{F: Temporally uncorrelated region}
In this region the temporal correlation in rupturing fibers vanishes. Since the spatial correlation still exists, the failure happens in a nucleating manner.


\subsection{Visualizing the failure modes}
Before we go to the description of the methods for drawing the phase boundaries, let us first look at the various failure modes described above. The
temporal configurations of the damages and stress profiles can give a qualitative idea of the different modes, which we will later describe in the
quantitative forms. 
\begin{figure}[tbh]
\centering
\includegraphics[width=8.5cm, keepaspectratio]{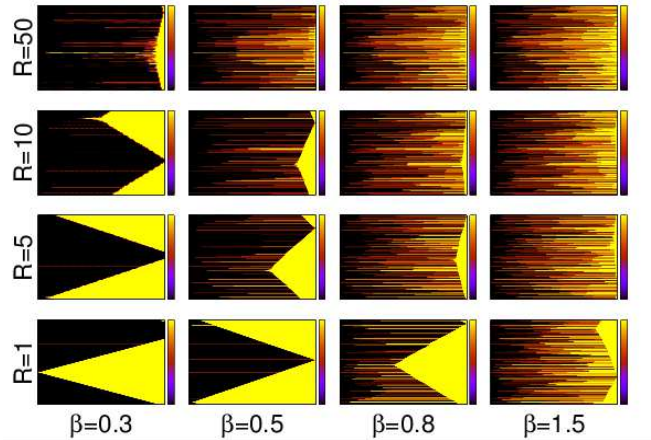}
\caption{The configurations of the failures in one dimension for different values of $R$ and $\beta$. The x-axis
is time and y-axis is the whole system. Zero stress imply broken fibers. For different values of the parameters
nucleation phenomenon can be clearly seen. The difference between the avalanche and percolative failures are not
apparent from the snap-shot, which will become clearer with the quantitative analysis in the following section. }
\label{1d_failures}
\end{figure}
In one dimension, it is easier to see the full temporal evolution of the damages and stress concentrations. In Fig. \ref{1d_failures}
 we plot the time evolution of the model for
different ranges of the $R, \beta$ parameters. The x-axis is the time, and in the y-axis the temporal stress profiles of the
system is shown, zero stress imply broken fibers. For low values of $\beta$ and $R$, we see clear nucleation, which eventually
engulfs the whole system. For slightly higher values, we see initial random failures, but in time a nucleation center grows, 
till the whole system collapses. For high values of $\beta$ and $R$, on the other hand, there is no nucleation, and the damage profile is
rather random in space. For this qualitative picture, it is not possible to see the distinctions between the temporally correlated failures for 
high $R$ and intermediate $\beta$ values, and the percolative failure for very high $\beta$ values. For that we need to look at the more quantitative 
measures described below. But this gives a pictorial sense of the damage profile and the dynamics prior to failure in the model for 
different ranges of values of $R$ and $\beta$. 

\begin{figure}[tbh]
\centering
\includegraphics[width=8.5cm, keepaspectratio]{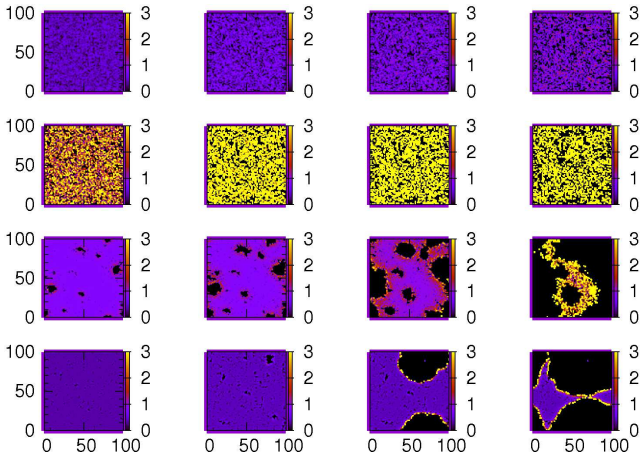}
\caption{The different modes of failures for two-dimensions are shown in terms of the stress profile at various times prior to failure
in a $100\times 100$ lattice, for different $R$ and $\beta$ values. The black regions are broken fibers. From top to bottom
the modes are avalanche, percolative, brittle and nucleating. Along horizontal axis snaps for different time steps are shown.
The times are not equispaced for different modes. In the avalanche mode ($\gamma=-1.0$, $\beta=0.6$), the time steps shown are 
415, 568, 630 and 676. For the 
percolating region ($\gamma=-6.0$, $\beta=2.5$), the steps are 83, 199, 269, 385. For the brittle region ($\gamma=-3.0$, $\beta=0.1$) 
99, 100, 101, 102. Finally, for the nucleation mode ($\gamma=-6.0$, $\beta=0.5$),
the time steps are 92, 165, 203 and 215. The stress profiles and damage configurations give a qualitative idea about the
different failure modes. For the avalanche mode in the top, there is no spatial correlation in damage and the stress profile is more
or less uniform. The similar feature can also be seen for the percolative failure, but in general with higher stress due to higher
disorder. For the brittle failure the stress is uniform too and the failure is very abrupt. For the nucleation, a stress
concentration in the spatially correlated damage region can clearly be seen.}
\label{2d_failures}
\end{figure}

In two dimensions, it is harder to see the temporal effects for obvious reasons. Nevertheless, in Fig. \ref{2d_failures} we plot the
stress/damage profile of the system for various modes of failures. The horizontal axis is snaps at different times.
Vertically from top to bottom we show the failures modes of avalanche, percolation, brittle and nucleation. It is
to be noted that the snaps are not in equal time intervals. In the avalanche process, we see that there is no spatial
correlation of the damages and the stress profiles are more or less uniform. In the percolation process too, there is
no spatial correlation in damage, but the stress values here goes to much higher values, since the disorder in very high
and there are many strong fibers. The principal distinction between the brittle region and the nucleation region is in
the time scales. While in the brittle region the snaps are unit time step apart, in the nucleation regions they are much 
further apart. It shows that in the brittle region the whole system collapse suddenly. On the other hand, in the 
nucleating region, the initial damages were random. But at later times one damaged area starts growing, due to the high stress
concentration at its boundary, which can also be clearly seen. This gives a qualitative idea about the phases of failures, 
which we will now discuss more quantitatively in terms of the phase boundaries.

\subsection{Description of the Phase Boundaries}

The various phases described above are separated by phase boundaries drawn on specific criteria. We will describe those now. 

\subsubsection{Quasi-brittle percolating (A) $-$ quasi-brittle nucleating (C) boundary}
\begin{figure}[tbh]
\centering
\includegraphics[width=8.5cm, keepaspectratio]{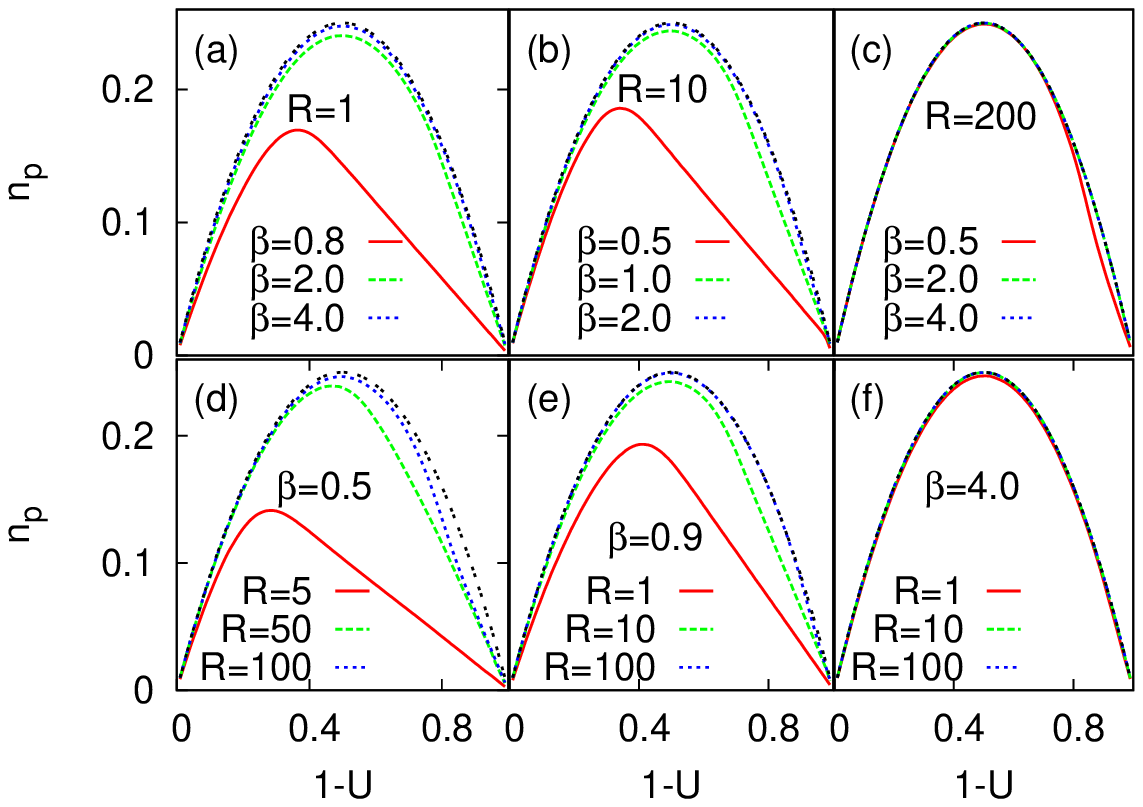} \\
\hspace{-0.33cm}\includegraphics[width=8.5cm, keepaspectratio]{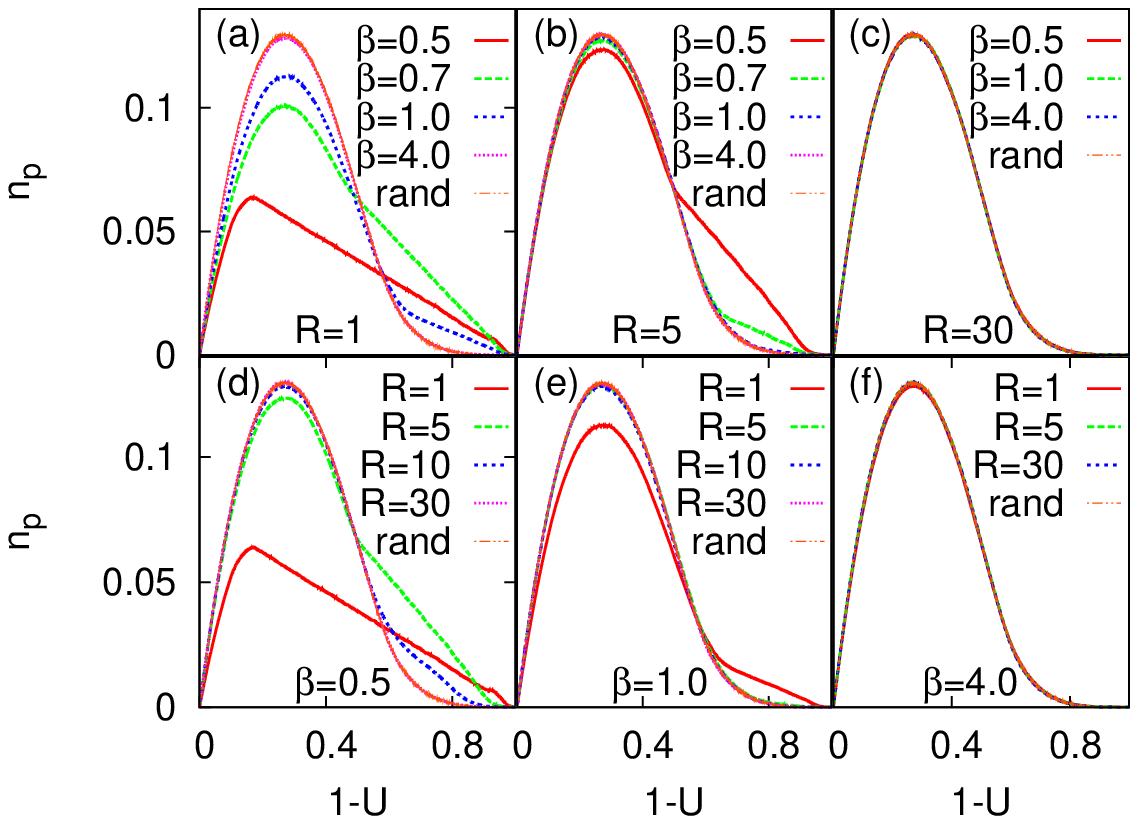}
\caption{The variations of number of patch per fiber ($n_p$) are shown for constant range ($R$) and different strength of disorder $\beta$ [(a-c) for one dimension and (d-f) for two dimension] and for constant strength of disorder and different ranges [(g-i) for one dimension and (j-l) for two dimension] with fraction of broken fibers $(1-U)$. It can be seen that for both high $R$ and high $\beta$ values, the curves merge with the ones obtained for completely random failures. The two limits, however, differ in term of dynamics, as discussed in the text.}
\label{patch}
\end{figure}

A general way to determine spatial correlation is to monitor the cluster density
with fraction of broken fibers. Fig. \ref{patch} shows the variation of cluster density $n_p$ (number of cluster divided by system size) with fraction of broken bonds $1-U$, at different $R$ and $\beta$ values, for both one and two dimension. In one dimension, the number of clusters of broken fibers is simply the number of side by side broken and unbroken fibers present. If $U$ is the fraction of surviving fibers at any time, then for complete random failure, the number of side by side broken and unbroken fiber will be $U(1-U)$ (normalized by system size). Any deviation  of $n_p$ from this function would  indicate spatial correlation.
A quantitative measure for such departure is the area under this $n_p$ v/s $(1-U)$ curve and compared it with the situation when the rupture is completely uncorrelated. In case of uncorrelated failure (for high $R$ or $\beta$) the area under the curve will be $A_{1d}=\displaystyle\int_{0}^{1}U(1-U)dU=1/6$. At low $R$ and $\beta$, the area under the curve deviates from $A_{1d}$. For two dimensions the situation is qualitatively similar. But the general shape of the curve for random failure is not known. However, there are many numerical studies in terms of random site percolation (see Ref. \cite{ziff} and references therein) that looks at density of patches under random occupations (see Fig.\ref{patch}).

\begin{figure}[ht]
\centering
\includegraphics[width=8.5cm, keepaspectratio]{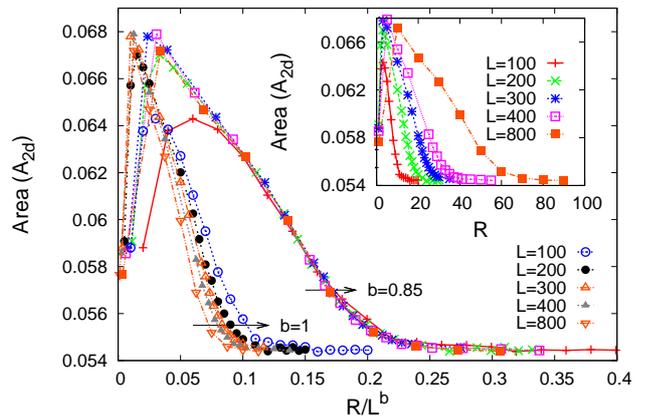}
\caption{For the two dimensional model, the scaling of the area under the patch density versus fraction of failed fibers curves (shown in Fig. 5) with R are shown. The linear scaling in the x-axis does not show satisfactory data collapse. The best collapse is seen when b = 0.85. The unscaled data are shown in the inset.}
\label{area}
\end{figure}

The deviation of the $n_p$ v/s $1-U$ curves from the random case determines this boundary. This gives a crossover scale $R_c$, which scales with the system size as $L^{2/3}$ \cite{pre15} in one-dimension. In two dimension the scaling changes to
\begin{equation}
 R_c\sim L^{b},
\label{rc_scaling}
\end{equation} 
with $b=0.85\pm 0.01$. Fig.\ref{area} shows the scaling of $R_c$ with system size $L$ in a two dimensional fiber bundle model. The areas ($A_{2d}$) under $n_p$ vs $1-U$ curves (see Fig.\ref{patch}) for different $L$ values are observed to scale with $RL^{-b}$, where $b=0.85$. 

One interesting implication of the scaling is, when the load sharing is a power law, the effective range of the load redistribution can be shown to be $R_{eff}\sim L^{3-\gamma}$, where $\gamma$ is the power of the load redistribution process. This can be understood through following calculation. With power law redistribution rule an effective range can be defined as:
\begin{equation}
 R_{eff}=\langle r\rangle=\displaystyle\int\limits_1^LrP(r)2\pi rdr=\displaystyle\frac{2-\gamma}{3-\gamma}\displaystyle\frac{L^{3-\gamma}-1}{L^{2-\gamma}-1},
\end{equation}
 where $P(r)\sim 1/r^{\gamma}$.
For $\gamma<2$, $R_{eff}\sim L$, implying mean-field regime. Also, for $\gamma>3$, $R_{eff}\sim const.$, therefore it is always local load sharing type. However, for $2<\gamma<3$, $R_{eff}\sim L^{3-\gamma}$ in the large system size limit. Since 
$R_c \sim L^b$, 
to get the crossover value for $\gamma$ we have to compare $R_{eff}(\gamma_c)\sim R_c$, giving 
\begin{equation}
\gamma_c=3-b. 
\label{gammac}
\end{equation}
But $b<1 (=0.85)$, giving $\gamma_c>2 (2.15)$. This explains an apparent result for $\gamma_c>2$ \cite{hidalgo}, which can now be claimed with much higher numerical accuracy.
\begin{figure}[ht]
\centering
\includegraphics[width=8.5cm, keepaspectratio]{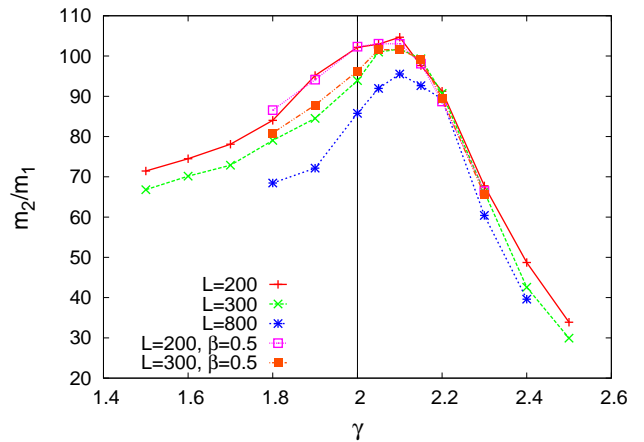}
\caption{The moment ratios of the cluster size distribution of the broken fibers in a two-dimensional power-law load sharing fiber bundle model are shown for different system sizes with $\gamma$. The peaks are consistently at $\gamma>2$. The threshold distributions are either uniform, or power-law with $\beta$ dependent cut-off, as mentioned.}
\label{moment_ratio}
\end{figure}
To verify this point, we have performed numerical simulations with power-law load redistribution. We have studied the cluster statistics of the broken fibers in the final stable configuration prior to complete failure (as was done in the paper by Hidalgo et al \cite{hidalgo}). We measured the moments of the cluster size distributions $n(s)$. The $k$-th moment is defined as 
\begin{equation}
m_k=\int s^kn(s)ds
\end{equation}
We have plotted the moment ratio $m_2/m_1$ in Fig.\ref{moment_ratio}. The peaks of the curves occur consistently above $\gamma=2$ for different system sizes and threshold distributions. 


\subsubsection{Brittle nucleating(B) $-$ brittle percolating (D) boundary}
The nature of this crossover line is the same as before and is drawn by monitoring the cluster density. The crossover length scale $R_c$ now scales non-universally with $L$, as $R_c\sim L^\zeta$ with $\zeta=\zeta(\beta)$ \cite{prep}. 


\subsubsection{High disorder nucleating (F) $-$ high disorder percolating (E) boundary}
This boundary is also drawn from the same measure as B-D boundary but the crossover scale here is independent of the strength of the disorder ($R_c \sim L^{2/3}$). 


\subsubsection{Brittle percolating (D) $-$ quasi-brittle percolating (C) boundary}
This class of boundaries separate brittle to quasi-brittle transitions. Particularly, in the brittle region, the breaking of the weakest fiber will cause the breakdown of the entire system. Hence, by measuring the fraction of surviving fibers in the last stable configuration before breakdown ($U_c$), we track the transition from brittle (with $U_c=1$) to quasi-brittle (with $U_c<1$) region.   A phase transition occurs only across this line \cite{epl15, chandreyee,Andersen,Silveira}, with no system size dependence of the transition line.


\subsubsection{Brittle nucleating (B) $-$ quasi-brittle nucleating (A) boundary}
This boundary is also drawn with the same criterion that leads to the D-C boundary. There is, however, a system size dependence of the line and it gets shifted to higher $\beta$ value with increasing system size in a inverse logarithmic manner (see Ref.\cite{ArxivSroy}), for a particular stress release range. 


\subsubsection{Quasi-brittle percolating (C) $-$ high disorder percolating (E) boundary}
\begin{figure}[tbh]
\centering
\includegraphics[width=8cm, keepaspectratio]{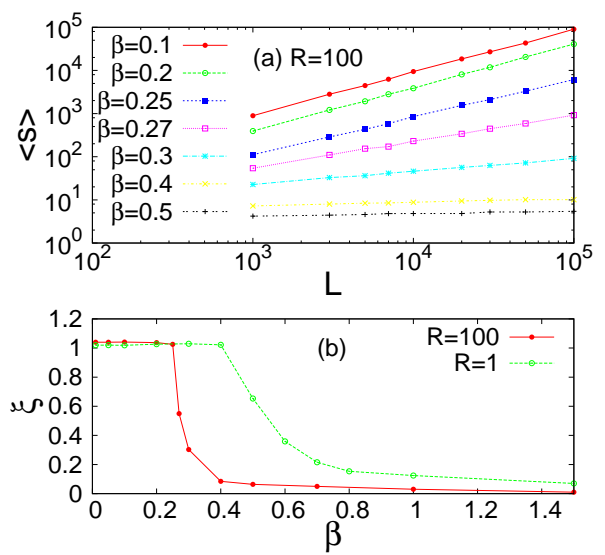}
\caption{(a) System size effect of $\langle s \rangle$ at various disorder values for for a particular stress release range (say $R=100$). $\langle s \rangle \sim L^{\xi}$, where $\xi$ is decreasing function of $\beta$. (b) $\xi$ as a function of disorder $\beta$.}
\label{Average_Avalanche_Lvariation}
\end{figure}

This boundary  separates the completely uncorrelated phase from temporally correlated quasi-brittle region C. We evaluate it in two different ways and the results match for the two cases. 
\begin{figure}[ht]
\centering
\includegraphics[height=5.5cm]{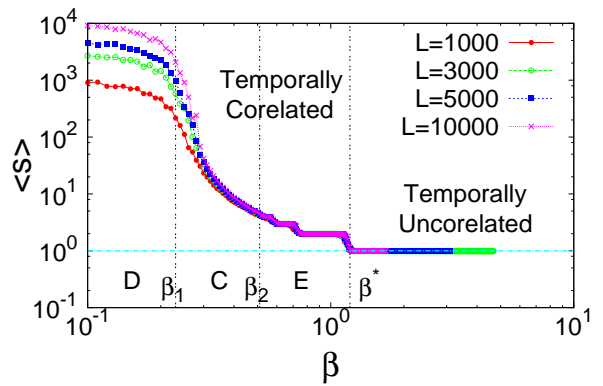}  
\caption{The variation of average avalanche size $\langle s \rangle$ with disorder for different $L$ values (in the mean-field limit). Middle: $\langle s \rangle/L$ vs $\beta$ for $10^3<L<10^4$.}
\label{Average_Avalanche_Betavariation}
\end{figure}
Fig.\ref{Average_Avalanche_Betavariation} shows the behavior of average avalanche size $\langle s \rangle$ in the mean filed limit against disorder $\beta$ and for system sizes ranging in between $10^3$ and $10^4$. We have also discussed the scaling of the average avalanche size $\langle s\rangle \sim L^{\xi}$ (see Fig.\ref{Average_Avalanche_Lvariation}). In the quasi-brittle region, $\xi$ is a (decreasing) function of $\beta$ and eventually $\langle s\rangle$ becomes independent of $L$ in the high disorder limit $E$. The $\beta$ value at which $\langle s\rangle$ becomes system size independent gives the boundary between C and E, because system size independence signifies complete removal of correlation in the system. 

\begin{figure}[ht]
\centering
\includegraphics[width=8.5cm]{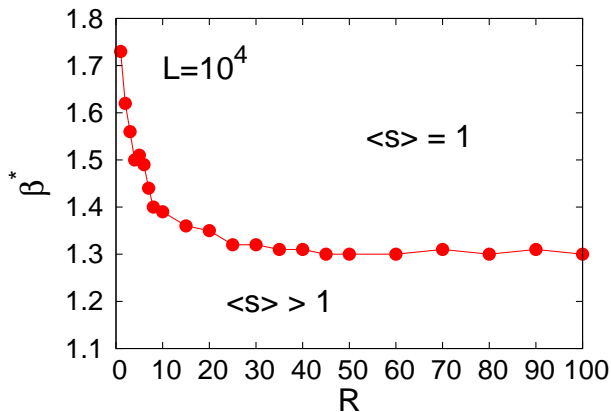}  
\caption{Behavior of $\beta^{\ast}$ with stress release range $R$. In the green region, there is no temporal correlation between the rupturing fibers.}
\label{BetaStar_Rvariation}
\end{figure}

Different regions, according to Fig.\ref{Average_Avalanche_Betavariation}, is described below (see Fig.\ref{Average_Avalanche_Lvariation} in support of the following behavior): 
\begin{enumerate}[label=(\roman*)]
\item For $\beta<\beta_1$, the failure process is brittle like abrupt. $\langle s \rangle \sim L$, since all the fibers break in a single avalanche.
\item For $\beta_1<\beta<\beta_2$, $\langle s \rangle \sim L^{\xi}$, where $\xi$ is an decreasing function of $\beta$ and reaches to a very low value at $\beta_2$ (shown in main text). In this region the bundle breaks in many avalanches like quasi-brittle material.
\item For $\beta_2<\beta<\beta^{\ast}$, $\langle s \rangle=k (>1)$. Here $k$ is function of $\beta$ only and independent of $L$. Very few avalanches are observed in this region.
\item The green vertical line (see Fig.\ref{Average_Avalanche_Betavariation}), drawn at $\beta=\beta^{\ast}$, shows an extreme limit of temporal correlation. The variation of $\beta^{\ast}$ with $R$ is shown in Fig.\ref{BetaStar_Rvariation}. With high local stress concentration (low $R$ value), $\beta^{\ast}$ is around 1.7. With increasing $R$, as the model entires the mean-field limit, $\beta^{\ast}$ saturates at 1.3. In the region $\beta>\beta^{\ast}$, the fibers break only by stress increment, giving $\langle s \rangle=1$.
\end{enumerate}
\begin{figure}[t]
\centering
\includegraphics[width=8.5cm, keepaspectratio]{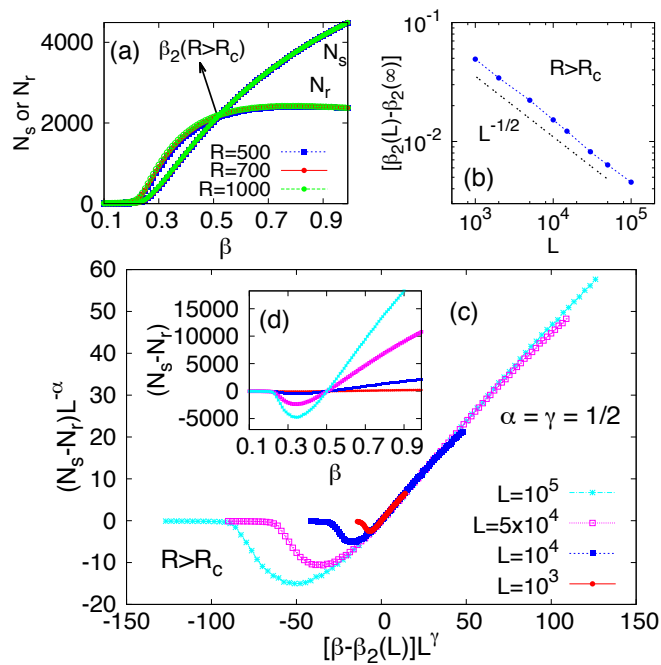}
\caption{(a) Variation of $N_s$ and $N_r$ as a function of $\beta$ for $L=10^4$. (b) System size effect of $\beta_2$, as the model approaches thermodynamic limit. (c) System size scaling of $(N_s-N_r)$ around $\beta=\beta_2$. The inset shows the unscaled behavior.}
\label{New_Figure1}
\end{figure}
A second way to approach the problem is to measure the number of stress increment $N_s$ and the number of times $N_r$ stress were redistributed during the entire time of survival of the system. Such interplay of $N_s$ and $N_r$ is shown in Fig.\ref{New_Figure1}. When $N_s$ outruns $N_r$, i.e. more fibers break due to stress increment (without spatial or temporal correlations) than due to stress redistributions, the uncorrelated region E is obtained. We find that the disorder strength $\beta_2$ when this happens scales with the system size as: $\beta_2(L)=\beta_2(\infty)+L^{-\alpha}$, with $\alpha=1/2$ (see Fig.\ref{New_Figure1}). Also, the system size scaling of $(N_s-N_r)$ is given by
\begin{equation}\label{Eq1}
(N_s-N_r)=L^{\alpha}\Phi[(\beta_2-\beta_2(L))L^{\gamma}],
\end{equation} 
with $\alpha=\gamma=1/2$. The $\beta_2$ obtained in this way matches with the boundary obtained from the scaling of the average avalanche size before.
Hence we conclude that $\beta_2$ is the range of disorder beyond which the system becomes completely uncorrelated (region E). 
\begin{figure}[ht]
\centering
\includegraphics[width=8.7cm]{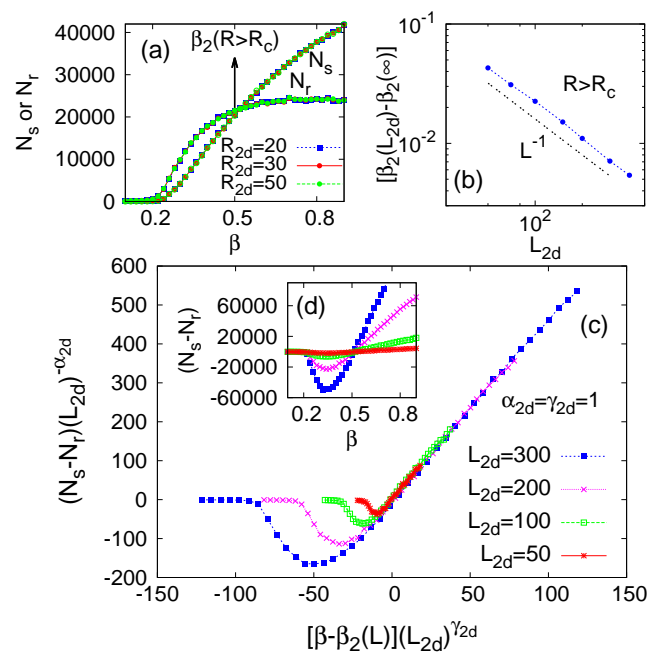}  
\caption{System size effect of $\beta_2$ for different stress release range, for two dimensional model. (a) Variation of $N_s$ and $N_r$ as a function of $\beta$ for $L=10^2$. (b) System size effect of $\beta_2$, as the model approaches thermodynamic limit. (c) System size scaling of $(N_s-N_r)$ around $\beta=\beta_2$. The inset shows the unscaled behavior.}
\label{New_Figure_2d}
\end{figure}

Fig.\ref{New_Figure_2d} shows this system size scaling of $(N_s-N_r)$ in the two dimension. The scaling shown in \ref{Eq1}, holds the same in two dimension also with respective exponents $\alpha_{2d}=\gamma_{2d}=1$. This in turn establish the scaling of $\beta_2$ as: $\beta_2(L)=\beta_2(\infty)+1/L$.


\subsubsection{Quasi-brittle nucleating (A) $-$ high disorder nucleating (F) boundary}
This boundary is drawn with the same criterion used to draw the boundary between C and E. Across this boundary the temporal correlation vanishes but the spatial correlation still exists. 
\begin{figure}[ht]
\centering
\includegraphics[width=8.0cm]{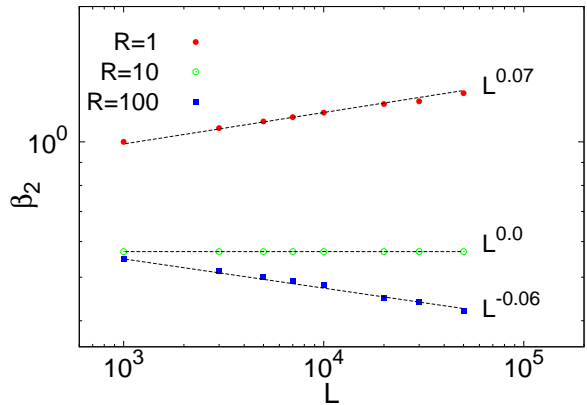}  
\caption{System size effect of $\beta_2$ for different stress release range $R$.}
\label{Beta2_Lvariation}
\end{figure}
At a certain stress release range $R$, $\beta_2$ is not being observed to change much while we alter the system size (see Fig.\ref{Beta2_Lvariation}). $\beta_2$ shows a scale free behavior with $L$ but with an extremely low exponent that suggest a very very weak system size dependence of $\beta_2$. \\  


Finally, using the criteria outlined above, we arrive at the quantitative phase diagram for fiber bundle model in one and two dimensions (see Fig.\ref{PD}). Almost all the studies in fiber bundle model fall in some point of this phase diagram. The most studied region being the region  $C$, which is also historically the earliest. Subsequently region $A$ was studied, which is qualitatively different from region $C$ in the sense that we no longer observe scale free avalanche statistics here. We provide a scaling criterion to separate these two regions. \\


\section{Discussions and conclusion}
In fracture of disordered solids, the two main factors determining the mode of failures are the range of interaction and the strength
of disorder in the solids. It is known that higher disorder produce higher precursory events \cite{vass} in a solid prior to failure, which is 
important in predicting catastrophic breakdowns. A transition from brittle to quasi-brittle modes of failure was observed
both theoretically  (see e.g. \cite{karpas11,epl15,papan17}) and experimentally (see e.g. \cite{Li92,sche10}) where 
the strength of disorder played a major role. However, the range of interaction i.e.
the region affected by the load redistribution following a local failure also plays a crucial role in determining the effect of
disorder strength. Generally, the compliance of the solid, determined by its elasticity, effect of agglomerate sizes, correlation or
plastic deformation etc. determines the range of interaction. A localized redistribution promotes stress/damage nucleation whereas 
the disorder promotes spreading of damage. It is the interplay between the two that gives many interesting effects in length and time
scales in various failure modes for fracture of disordered solids. In this work we have addressed the interplay of these two effects
using a random fiber bundle model in one and two dimensions. In isolation some of the limiting cases were studied before. But the
full range of localization of strength and width of the disorder distribution give various phases and boundaries across which the relative
influence of these two competing effects vary. In particular, we recover by increasing the range of interaction, a region with no spatial
correlation, where the temporal correlation still exists (avalanche region C) that survives in the large system size limit (see Eq. \ref{rc_scaling}), which
was absent in the random fuse model. In that model the eventual nucleation was always dominant in the large system size limit (equivalent of 
region A). Experimentally, of course such regions are observed for many decades (see \cite{books} for detailed discussions). Furthermore, we are also able to verify the
unusual scaling of the interaction range that leads to nucleation. The fact that $\gamma_c>2$ in Eq. \ref{gammac} ($b<1$) has interesting consequences particularly for
fracture, given the inverse square interaction is usually expected for elastic solids. The criteria for drawing different phase boundaries and the size-scaling
in each of those phases are summarized in forms of table in the Appendix. It is also to be noted that the phase boundaries sometimes have dependence
on system size (see e.g. Eq. (\ref{rc_scaling})), therefore appropriate finite size scaling needs to be done (as are mentioned for applicable cases) in order to translate the results into
different system sizes.
 
There have been many studies over the years in interpolating between various phases of the fiber bundle model described above. Among these, most efforts were concentrated in interpolating between regions $A$ and  $C$, because this region gives the critical interaction range below which the eventual failure will be nucleation dominated, a much debated topic in fracture \cite{zapperi97}. The crossover from $A$ to $C$ was also accessed in Ref. \cite{stormo12} by tuning the elastic modulus of the bottom plate of the model, which in turn controls the range of interaction. In Ref. \cite{epl15} the authors moved from region $D$ to $C$ in the mean-field limit. The value of $\beta$ was exactly calculated in the mean-field limit \cite{chandreyee}. Similar transitions between D and C phases for a generic class of disorder distribution was also noted in refs.\cite{Andersen,Silveira}.
 
Many experimental observations, like brittle (region $B$ \& $D$) to quasi-brittle/ductile (region $A$ \& $C$) transition \cite{Li92}, scale free
size distribution for acoustic emission \cite{guarino98}, subsequent damage nucleation \cite{guarino98} etc., can also be explained by this phase diagram. Such properties are characteristic of region $A$, where the so called `finite size criticality' \cite{Shekhawat} is observed, i.e. the system starts off giving scale free avalanches, but the final failure is nucleation driven. Unlike random resistor network \cite{Moreira}, where nucleation always dominates in the final failure mode, in the fiber bundle model phase diagram, there exists a temporally correlated failure mode that sustains in the thermodynamics limit.

In conclusion, we have obtained a phase diagram for failure of disordered solids using random fiber bundle model. We describe all distinct modes of failure with varying disorder ($\beta$) and stress release range ($R$). Disorder effects the abruptness of the failure process while the stress release range influences the correlation between successive rupturing of fibers. Interplay of these two affects leads to spatial and/or temporal correlation or random failures.  The resulting phase diagram gives a framework for understanding previous theoretical and experimental attempts to interpolate between these different failure modes. 

SB acknowledges Alexander von Humboldt foundation for funding.


\twocolumngrid



\onecolumngrid
\begin{flushleft} \textbf{\textit{\Large{Appendix}}} \end{flushleft}

Below we have shown all the phases observed in the model in a tabular form. The properties of individual phases as well as the criterion for drawing the phase boundaries are presented below : \\ \\

\textbf{\underline{Properties of Different Regions}:}

\begin{table}[h]
\begin{tabular}{|c|c|c|c|c|}
 \hline
 $Region$ & \multicolumn{2}{ c }{\hspace{5cm} Properties} & \multicolumn{2}{ c| }{} \\ \cline{2-5}
 & Abruptness & Failure Pattern & Average Avalanche Size $\langle s \rangle$ & $N_s$ v/s $N_r$ \\ \hline
 $B$ & Abrupt failure: $U_c\approx1$ & Nucleating & $\langle s \rangle \sim L$ & $N_s=1,N_r>N_s$ \\ \hline
 $D$ & Abrupt failure: $U_c\approx1$ & Percolating & $\langle s \rangle \sim L$ & $N_s=1,N_r>N_s$ \\ \hline
 $C$ & Non-abrupt failure: $U_c<1$ & Percolating & $\langle s \rangle \sim L^{\xi}$: $\xi$ decreases with $\beta$ & $N_s>1,N_r>N_s$ \\ \hline
 $A$ & Non-abrupt failure: $U_c<1$ & Nucleating & $\langle s \rangle \sim L^{\xi}$: $\xi$ decreases with $\beta$ & $N_s>1,N_r>N_s$ \\ \hline
 $E$ & Non-abrupt failure: $U_c$ is very low & Percolating & $\langle s \rangle=Constant \ (>1)$ & $N_s>N_r$ \\ \hline
 $F$ & Non-abrupt failure: $U_c$ is very low & Nucleating & $\langle s \rangle=Constant \ (>1)$ & $N_s>N_r$ \\ \hline
\end{tabular}
\end{table}

\textbf{\underline{Criteria for Drawing the Boundaries}:}

\begin{table}[ht]
\begin{tabular}{|c|c|}
\hline
Boundary & Criteria \\ \hline
B-D or A-C or F-E & The area $A$ under $n_p$ v/s $(1-U)$ plot approaches the area $A_{rand}$ for random failure: $(A_{rand}-A)<10^{-4}$. \\ \hline
D-C or B-A & Critical fraction of unbroken bonds deviates form 1: $U_c<1$.  \\ \hline
C-E or A-F & $\xi$ reaches zero and $\langle s \rangle$ becomes independent of $L$. Also where $N_s$ outruns $N_r$.  \\ \hline 
\end{tabular}
\end{table}


\end{document}